\documentclass[%
aip,
jcp,
 amsmath,amssymb,
reprint,%
]{revtex4-1}

\usepackage{graphicx}
\usepackage{dcolumn}
\usepackage{bm}

\usepackage[utf8]{inputenc}
\usepackage[T1]{fontenc}
\usepackage{mathptmx}
\usepackage{physics}
\usepackage{ytableau}
\usepackage{amsmath}

\allowdisplaybreaks

\DeclareSymbolFont{newfont}{OML}{cmm}{m}{it}
\DeclareMathSymbol{\epsilon}{3}{newfont}{15}
\DeclareMathSymbol{\varrho}{3}{newfont}{37}

\begin{document}

\preprint{AIP/123-QED}

\title[CARS in optically active medium]{Coherent Anti-Stokes Raman scattering in optically active medium}

\author{Tuguldur Kh. Begzjav}
\affiliation{Institute for Quantum Science and Engineering, Department of Physics and Astronomy,\\ Texas A\&M University, College Station, TX 77843, USA}
 \email{mn.tuguldur@tamu.edu}

\author{Marlan O. Scully}
\affiliation{Institute for Quantum Science and Engineering, Department of Physics and Astronomy,\\ Texas A\&M University, College Station, TX 77843, USA}
\affiliation{Department of Physics, Baylor University, Waco, TX 76706, USA}
\affiliation{Department of Mechanical and Aerospace Engineering, Princeton University, Princeton, NJ 08544, USA}

\author{Girish S. Agarwal}%
\affiliation{Institute for Quantum Science and Engineering, Department of Physics and Astronomy,\\ Texas A\&M University, College Station, TX 77843, USA}
\affiliation{Department of Biological and Agricultural Engineering, Texas A\&M University, College Station, TX 77843, USA}

\date{\today}

\begin{abstract}
Early theoretical works on coherent anti-Stokes Raman scattering in optically active medium consider only heterodyne signal and subsequently, fourth- and fifth-rank tensor averages have been used. In this work, we presented a full signal expression of coherent anti-Stokes Raman scattering in optically active medium with the help of eighth- and ninth-rank tensor averaging for simplest experimental configuration namely, measurements of post-selected circularly polarized components of scattered anti-Stokes field in the presence of three incident laser beams all linearly polarized along the same axis.
\end{abstract}

\maketitle

\section{Introduction}
One of the most ubiquitous tools for molecular chiral study is spectroscopic tool called vibrational optical activity including vibrational circular dichroism\cite{Nafie1976} and Raman optical activity\cite{Barron1971,Barron1973,Hecht1991}. This type of spectroscopic tool has been well studied and developed last half-century. Almost simultaneously, high-order nonlinear coherent spectroscopic techniques came into mainstream and have been considered for enhanced optically active signals\cite{Begzjav2019coherence}. Primarily, there are two candidates for nonlinear spectroscopic techniques: a) Coherent anti-Stokes Raman scattering (CARS) in optically active medium \cite{Bjarnason1980,Oudar1982} b) coherent five-wave mixing in optically active medium known as BioCARS\cite{Koroteev1995,Zheltikov1999}.
Nowadays, experimental realizations for these nonlinear techniques are still challenging for experimentalists especially in the case of BioCARS.
However, some recent progress on CARS technique in optically active medium is made by K.~Hiramatsu et al\cite{Hiramatsu2012,Hiramatsu2013,Hiramatsu2015}. They have tested various types of CARS spectroscopic tools for chiral discrimination and reported two-orders of higher signal strength than spontaneous Raman optical activity signal.

Early theoretical works\cite{Bjarnason1980,Oudar1982} on CARS in optically active medium based on rotational averaging of third order nonlinear susceptibility are only valid for heterodyne detection scheme. 
Here, motivated by these works, we presented a rigorous and complete theory for CARS signal in optically active medium that consists of randomly oriented molecules. Our theory is based on isotropic rotational averaging of CARS signal strength itself rather than susceptibility. This generalization of the theory leads to high-rank tensor averaging.

Mainly, there are two basic ways of measurement of spectroscopic signal: a) post-selected polarization measurement where specifically chosen polarization component of scattered signal is measured b) full signal measurement where all polarization components of the scattered signal are measured. Obviously, these two ways of measurement result in tensors of different ranks. For example, post-selected polarization measurement requires the rotational average of eighth- and ninth-rank tensors whereas full signal measurement needs the rotational average of only sixth- and seventh-rank tensors.
Recent developments on rotational averaging of high-rank tensors\cite{Nessler2019,Begzjav2019} along with the seminal works by D.~L.~Andrews et al.\cite{Andrews1977,Andrews1981,Andrews1982,Andrews1989} enable us to find rotational average values of nonlinear spectroscopic post-selected signals from randomly oriented chiral molecules.

\section{Quantum electrodynamic theory of CARS}
In our theoretical model, three monochromatic laser beams with wave vectors $\mathbf{k}_1$, $\mathbf{k}_2$ and $\mathbf{k}_3$, and polarizations $\mathbf{e}^{(1)}$, $\mathbf{e}^{(2)}$ and $\mathbf{e}^{(3)}$ impinge upon sample, and coherently scattered beam of wave vector $\mathbf{k}_4$ and polarization $\mathbf{e}^{(4)}$ is produced.  
Hamiltonian of this matter--field system is written by \cite{Craig1998}
\begin{align}
\hat{H}=\hat{H}_0+\hat{H}_{\text{int}},
\end{align}
where
$\hat{H}_0$ is free field and matter Hamiltonian, and
\begin{align}\label{intHam}
\hat{H}_{\text{int}}=&
\sum_{j=1}^4
\left[
-i\sqrt{\frac{\hbar ck_j}{2\varepsilon_0 V}} \left(\mathbf{e}^{(j)} \hat{a}_j -\bar{\mathbf{e}}^{(j)} \hat{a}_j^\dagger\right)
\cdot\hat{\pmb{\mu}}\right.\nonumber\\
&-
i\sqrt{\frac{\hbar k_j}{2\varepsilon_0 c V}}\left(\hat{\mathbf{k}}_j\times\mathbf{e}^{(j)} \hat{a}_j - \hat{\mathbf{k}}_j\times\bar{\mathbf{e}}^{(j)} \hat{a}_j^\dagger\right)\cdot \hat{\mathbf{m}}\nonumber\\
&
\left.+\frac{1}{3}\sqrt{\frac{\hbar ck_j}{2\varepsilon_0 V}}
\sum_{\alpha,\beta}\hat{q}_{\alpha\beta} k_{j,\alpha} 
\left(e_\beta^{(j)} \hat{a}_j +\bar{e}_\beta^{(j)} \hat{a}_j^\dagger\right)
\right]
\end{align}
is interaction Hamiltonian. Here, $V$ is quantization volume, $\varepsilon_0$ is vacuum permittivity and $c$ is speed of light in vacuum. Annihilation and creation operators of the $j$th laser beam are denoted by $\hat{a}_j$ and $\hat{a}^\dagger_j$, respectively. Here, we take not only electric dipole interaction but also magnetic dipole and electric quadrupole interactions. 
Electric dipole, magnetic dipole and electric quadrupole operators are denoted by $\hat{\pmb{\mu}}$, $\hat{\mathbf{m}}$ and $\hat{\mathbf{q}}$, respectively. Greek letters $\alpha$ and $\beta$ in $\hat{q}_{\alpha\beta}$ represent Cartesian components $\{x,y,z\}$. The vectors $\hat{\mathbf{k}}_i$ with hat represent unit vectors along the vectors $\mathbf{k}_i$.

\begin{figure*}
\includegraphics[scale=0.85]{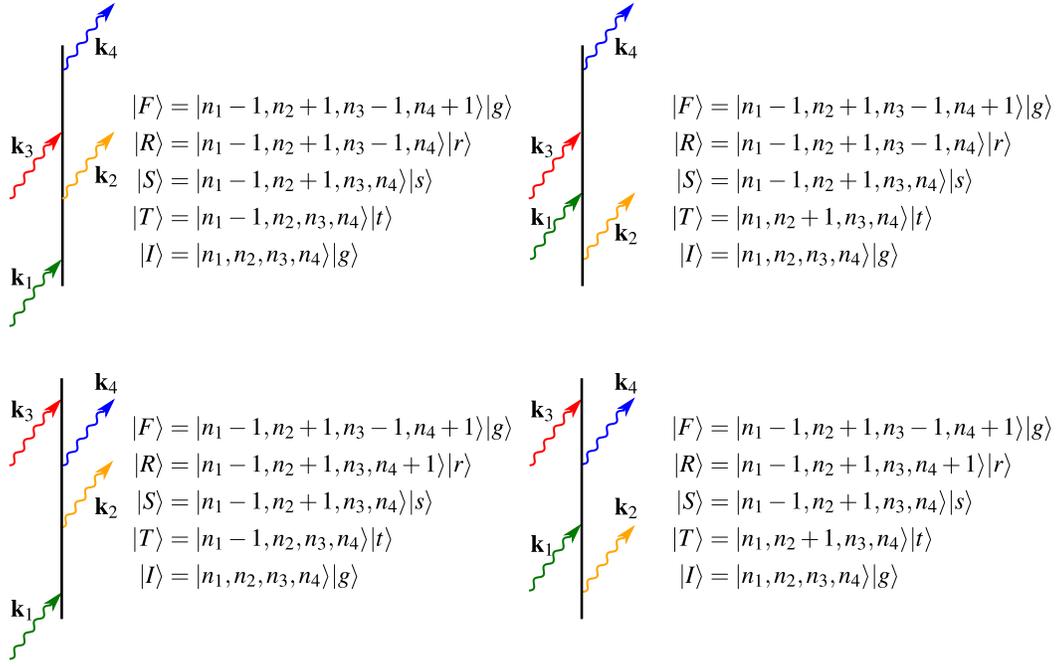}
\caption{Non-relativistic Feynman diagrams associated with four different time ordering of CARS process. Labels represent of wave vectors of each beam. On the right of each Feynman diagram, initial, final and intermediate states of the system are shown. The quantum states with letters in lowercase represent molecular quantum states whereas the quantum states with letters in uppercase represent quantum states of entire system.}\label{fig1}
\end{figure*}

Non-relativistic Feynman diagrams for CARS processes are shown in Fig.~\ref{fig1}.
Scattering matrix $M_{FI}$ is given by
\begin{align}
M_{FI}=\sum_{R,S,T}
&\langle F\vert \hat{H}_{\text{int}}
\vert R\rangle\langle R\vert
\frac{1}{E-\hat{H}_0+i\epsilon}
\vert R\rangle\nonumber\\
&\langle R\vert
\hat{H}_{\text{int}}
\vert S\rangle
\langle S\vert
\frac{1}{E-\hat{H}_0+i\epsilon}
\vert S\rangle\nonumber\\
&
\langle S\vert
\hat{H}_{\text{int}}
\vert T\rangle\langle T\vert
\frac{1}{E-\hat{H}_0+i\epsilon}
\vert T\rangle\langle T\vert
\hat{H}_{int}\vert I\rangle,
\end{align}
where $E=n_1\hbar\omega_1+n_2\hbar\omega_2+n_3\hbar\omega_3+n_4\hbar\omega_4+E_g$ is total energy of the system and $\epsilon$ is a positive number. Here, number of photons in the $j$th laser beam is denoted by $n_j$, and molecular energy of ground state $b$ is denoted by $E_g$. Initial state $\vert I\rangle$, intermediate states $\vert T\rangle$, $\vert S\rangle$ and $\vert R\rangle$, and final state $\vert F\rangle$ for each Feynman diagram are defined in the Fig.~\ref{fig1}. 
When we take only electric dipole interaction in the Hamiltonian~\ref{intHam}, the scattering matrix $M_{FI}^{\pmb{\mu}}$ is obtained as follows \cite{Mukamelbook}
\begin{align}\label{electric_part}
M_{FI}^{\pmb{\mu}}
=&
-i\pi\rho_s
\left(
\frac{\hbar c}{2\varepsilon_0 V}
\right)^2
\sqrt{k_1k_2k_3k_4}\sqrt{n_1n_3(n_2+1)(n_4+1)}\nonumber\\
&\times
\bar{e}^{(4)}_i e^{(3)}_j \bar{e}^{(2)}_k e^{(1)}_l
\alpha_{ij}(-\omega_3,\omega_4)
\alpha_{kl}(-\omega_1,\omega_2),
\end{align}
where $\rho_s$ is density of states in Stokes laser beam, and
\begin{align}\label{poltensor1}
\alpha_{ij}(-\omega_3,\omega_4)
=
\sum_{r}
\left(
\frac{
\langle f\vert\hat{\mu}_i\vert r\rangle
\langle r\vert\hat{\mu}_j\vert s\rangle
}{[E_{rs} - \hbar\omega_3-i\epsilon]}
+
\frac{
\langle f\vert\hat{\mu}_j\vert r\rangle
\langle r\vert\hat{\mu}_i\vert s\rangle
}{[E_{rs} + \hbar\omega_4+i\epsilon]}
\right),
\end{align}
and
\begin{align}\label{poltensor2}
\alpha_{kl}(-\omega_1,\omega_2)
=
\sum_{t}
\left(
\frac{\langle s\vert\hat{\mu}_k\vert t\rangle
\langle t\vert\hat{\mu}_l\vert g\rangle}{[E_{tg}-\hbar \omega_1-i\epsilon]}
+
\frac{\langle s\vert\hat{\mu}_l\vert t\rangle
\langle t\vert\hat{\mu}_k\vert g\rangle}{[E_{tg}+\hbar \omega_2+i\epsilon]}
\right)
\end{align}
are electric dipole polarizability tensors. Accordingly, the scattering matrix $M_{FI}^{(\mathbf{m},\mathbf{q})}$ associated with magnetic dipole and electric quadrupole interactions is obtained as
\begin{align}\label{m_and_q}
M_{FI}^{(\mathbf{m},\mathbf{q})}&=
-i\pi\rho_s
\left(
\frac{\hbar c}{2\varepsilon_0 V}
\right)^2
\sqrt{k_1k_2k_3k_4}\sqrt{n_1n_3(n_2+1)(n_4+1)}\nonumber\\
&
\left(
\frac{1}{c}
\bar{e}^{(4)}_i e^{(3)}_j
\bar{e}^{(2)}_k 
(\hat{\mathbf{k}}_1 \times
\mathbf{e}^{(1)})_l
\alpha_{ij}(-\omega_3,\omega_4)
G_{kl}^{(1)}(-\omega_1,\omega_2)\right.\nonumber\\
&+
\frac{1}{c}
\bar{e}^{(4)}_i e^{(3)}_j
(\hat{\mathbf{k}}_2\times
\bar{\mathbf{e}}^{(2)} ) _k 
e^{(1)}_l
\alpha_{ij}(-\omega_3,\omega_4)
G_{kl}^{(2)}(-\omega_1,\omega_2)\nonumber\\
&+
\frac{1}{c}
\bar{e}^{(4)}_i 
(\hat{\mathbf{k}}_3\times
e^{(3)})_j
\bar{e}^{(2)}_k 
e^{(1)}_l
G_{ij}^{(1)}(-\omega_3,\omega_4)
\alpha_{kl}(-\omega_1,\omega_2)\nonumber\\
&+
\frac{1}{c}
(\hat{\mathbf{k}}_4\times
\bar{e}^{(4)})_i 
e^{(3)}_j
\bar{e}^{(2)}_k 
e^{(1)}_l
G_{ij}^{(2)}(-\omega_3,\omega_4)
\alpha_{kl}(-\omega_1,\omega_2)\nonumber\\
&+
\frac{i}{3}
\bar{e}^{(4)}_i e^{(3)}_j \bar{e}^{(2)}_k 
e^{(1)}_l  k_{1,n}
\alpha_{ij}(-\omega_3,\omega_4)
A^{(1)}_{k,ln}(-\omega_1,\omega_2)\nonumber\\
&-
\frac{i}{3}
\bar{e}^{(4)}_i e^{(3)}_j 
\bar{e}^{(2)}_k
k_{2,n} 
e^{(1)}_l 
\alpha_{ij}(-\omega_3,\omega_4)
A^{(2)}_{l,kn}(-\omega_1,\omega_2)
\nonumber\\
&+
\frac{i}{3}
\bar{e}^{(4)}_i 
e^{(3)}_j k_{3,n}
 \bar{e}^{(2)}_k e^{(1)}_l
A^{(1)}_{i,jn}(-\omega_3,\omega_4)
\alpha_{kl}(-\omega_1,\omega_2)\nonumber\\
&-
\left.
\frac{i}{3}
\bar{e}^{(4)}_i
k_{4,n}
e^{(3)}_j \bar{e}^{(2)}_k e^{(1)}_l
A^{(2)}_{j,in}(-\omega_3,\omega_4)
\alpha_{kl}(-\omega_1,\omega_2)
\right)
\end{align}
where $G_{kl}^{(1)}(-\omega_1,\omega_2)$, $G_{kl}^{(2)}(-\omega_1,\omega_2)$, $G_{ij}^{(1)}(-\omega_3,\omega_4)$ and $G_{ij}^{(2)}(-\omega_3,\omega_4)$ are electric dipole--magnetic dipole optical activity tensors and $A^{(1)}_{k,ln}(-\omega_1,\omega_2)$, $A^{(2)}_{l,kn}(-\omega_1,\omega_2)$, $A^{(1)}_{i,jn}(-\omega_3,\omega_4)$ and $A^{(2)}_{j,in}(-\omega_3,\omega_4)$ are electric dipole--electric quadrupole optical activity tensors. The explicit forms of these tensors are given in Appendix~\ref{app_a}.

Mathematical properties of the tensors in Eqs.~\ref{electric_part},\ref{m_and_q} are considerably useful further calculations. The first property that much helps on calculation is involved with dipole and quadrupole moments. Particularly, it is well known that electric dipole and quadrupole operators (magnetic dipole operator) can be considered as purely real (imaginary) on the basis of molecular wavefunctions unless there is an external magnetic field. Secondly, according to Born-Oppenheimer approximation molecular wavefunctions can be a direct product of electronic and vibrational wavefunctions. With the help of above two arguments and neglecting small contribution due to nuclear motion, firstly the tensor $\alpha_{ij}$ becomes symmetric real-valued and secondly four optical activity tensors reduce to two real-valued tensors namely $G'_{ij}$ and $A_{i,jn}$ defined as
\begin{align}
G'_{ij}&=iG_{ij}^{(1)}=-iG_{ji}^{(2)},\nonumber\\
A_{i,jn}&=A_{i,jn}^{(1)}=A_{i,jn}^{(2)}.
\end{align}
Meantime, we ignore the small number $\epsilon$ in the denominator of polarizability tensor $\alpha_{ij}$ and optical activity tensors $G'_{ij}$ and $A_{i,jn}$ assuming the corresponding transitions are far from a resonance. 
It is also important to note that the tensor $A_{i,jn}$ is symmetric under permutation of indices $j$ and $n$ due to symmetric nature of electric quadrupole tensor operator\cite{Jackson1975} $\hat{q}_{\alpha\beta}$.

Exploiting the real-valued nature of the tensors $\alpha_{ij}$, $G'_{ij}$ and $A_{i,jn}$ we obtained simplified expression for
total scattering matrix $M_{FI}$ that is simply sum of Eq.~\ref{electric_part} and Eq.~\ref{m_and_q}. Its explicit form is
\begin{widetext}
\begin{align}\label{general}
&\vert M_{FI}\vert^2
=\pi^2\rho_s^2
\left(
\frac{\hbar c}{2\varepsilon_0 V}
\right)^4
k_1k_2k_3k_4 n_1n_3(n_2+1)(n_4+1)\nonumber\\
&\left(
e^{(4)}_m \bar{e}^{(3)}_o e^{(2)}_p \bar{e}^{(1)}_q
\alpha_{mo}(-\omega_3,\omega_4)
\alpha_{pq}(-\omega_1,\omega_2)
\bar{e}^{(4)}_i e^{(3)}_j \bar{e}^{(2)}_k e^{(1)}_l
\alpha_{ij}(-\omega_3,\omega_4)
\alpha_{kl}(-\omega_1,\omega_2)\right.\nonumber\\
&-
\left.
\frac{2}{c}
\Im{
e^{(4)}_m \bar{e}^{(3)}_o
e^{(2)}_p
(\hat{\mathbf{k}}_1 \times
\mathbf{\bar{e}}^{(1)})_q
\bar{e}^{(4)}_i e^{(3)}_j \bar{e}^{(2)}_k e^{(1)}_l
}
\alpha_{mo}(-\omega_3,\omega_4)
G_{pq}'(-\omega_1,\omega_2)
\alpha_{ij}(-\omega_3,\omega_4)
\alpha_{kl}(-\omega_1,\omega_2)\right.\nonumber\\
&+
\left.
\frac{2}{c}
\Im{
e^{(4)}_m \bar{e}^{(3)}_o
(\hat{\mathbf{k}}_2\times
\mathbf{e}^{(2)} ) _p 
\bar{e}^{(1)}_q
\bar{e}^{(4)}_i e^{(3)}_j \bar{e}^{(2)}_k e^{(1)}_l
}
\alpha_{mo}(-\omega_3,\omega_4)
G_{qp}'(-\omega_1,\omega_2)
\alpha_{ij}(-\omega_3,\omega_4)
\alpha_{kl}(-\omega_1,\omega_2)\right.\nonumber\\
&-
\left.
\frac{2}{c}
\Im{
e^{(4)}_m 
(\hat{\mathbf{k}}_3\times
\bar{e}^{(3)})_o
e^{(2)}_p 
\bar{e}^{(1)}_q
\bar{e}^{(4)}_i e^{(3)}_j \bar{e}^{(2)}_k e^{(1)}_l
}
G_{mo}'(-\omega_3,\omega_4)
\alpha_{pq}(-\omega_1,\omega_2)
\alpha_{ij}(-\omega_3,\omega_4)
\alpha_{kl}(-\omega_1,\omega_2)\right.\nonumber\\
&+
\left.
\frac{2}{c}
\Im{
(\hat{\mathbf{k}}_4\times
e^{(4)})_m 
\bar{e}^{(3)}_o
e^{(2)}_p 
\bar{e}^{(1)}_q
\bar{e}^{(4)}_i e^{(3)}_j \bar{e}^{(2)}_k e^{(1)}_l
}
G_{om}'(-\omega_3,\omega_4)
\alpha_{pq}(-\omega_1,\omega_2)
\alpha_{ij}(-\omega_3,\omega_4)
\alpha_{kl}(-\omega_1,\omega_2)\right.\nonumber\\
&+
\left.
\frac{2}{3}
\Im{
e^{(4)}_m \bar{e}^{(3)}_o e^{(2)}_p 
\bar{e}^{(1)}_q  k_{1,n}
\bar{e}^{(4)}_i e^{(3)}_j \bar{e}^{(2)}_k e^{(1)}_l
}
\alpha_{mo}(-\omega_3,\omega_4)
A_{p,qn}(-\omega_1,\omega_2)
\alpha_{ij}(-\omega_3,\omega_4)
\alpha_{kl}(-\omega_1,\omega_2)\right.\nonumber\\
&-
\left.
\frac{2}{3}
\Im{
e^{(4)}_m \bar{e}^{(3)}_o 
e^{(2)}_p
k_{2,n} 
\bar{e}^{(1)}_q 
\bar{e}^{(4)}_i e^{(3)}_j \bar{e}^{(2)}_k e^{(1)}_l
}
\alpha_{mo}(-\omega_3,\omega_4)
A_{q,pn}(-\omega_1,\omega_2)
\alpha_{ij}(-\omega_3,\omega_4)
\alpha_{kl}(-\omega_1,\omega_2)\right.\nonumber\\
&+
\left.
\frac{2}{3}
\Im{
e^{(4)}_m 
\bar{e}^{(3)}_o k_{3,n}
e^{(2)}_p \bar{e}^{(1)}_q
\bar{e}^{(4)}_i e^{(3)}_j \bar{e}^{(2)}_k e^{(1)}_l
}
A_{m,on}(-\omega_3,\omega_4)
\alpha_{pq}(-\omega_1,\omega_2)
\alpha_{ij}(-\omega_3,\omega_4)
\alpha_{kl}(-\omega_1,\omega_2)\right.\nonumber\\
&-
\left.
\frac{2}{3}
\Im{
e^{(4)}_m
k_{4,n}
\bar{e}^{(3)}_o e^{(2)}_p \bar{e}^{(1)}_q
\bar{e}^{(4)}_i e^{(3)}_j \bar{e}^{(2)}_k e^{(1)}_l
}
A_{o,mn}(-\omega_3,\omega_4)
\alpha_{pq}(-\omega_1,\omega_2)
\alpha_{ij}(-\omega_3,\omega_4)
\alpha_{kl}(-\omega_1,\omega_2)\right).
\end{align}
\end{widetext}
According to Eq.~\ref{general}, if all polarization vectors from $\mathbf{e}_1$ to $\mathbf{e}_4$ are purely real vectors, then no optically active signal is generated at all. This is the similar situation as linear polarization Raman optical activity\cite{Hecht1990} proposed by L.~Hecht et al. They claim that optical activity tensors must have nonzero imaginary parts to produce chiral signal when incident and scattered fields are both linearly polarized i.e. polarization vectors are purely real.
This is an analogous argument that we observe from Eq.~\ref{general}. 

Accessible experimental quantity for CARS can be transition rate $T_{CARS}$. In general, it is calculated using Fermi's golden rule\cite{Craig1998} taking quantum scattering matrix $M_{FI}$ into account as follows
\begin{align}\label{goldenrule}
T_{CARS}=\frac{2\pi}{\hbar}\rho_f\vert M_{FI}\vert^2
\end{align}
where $\rho_f$ is a density of states in anti-Stokes mode. In the next section, we refine our general result \ref{general} in the case of simplest polarization configuration of the input and scattered fields.

\section{Rotational average of CARS signal}
It is still well common that the most of modern spectroscopic tools rely on bulk samples rather than a single molecule. Meantime, randomly oriented molecules in bulk sample provide different responses under influence of external laser fields. Therefore, three-dimensional rotational averaging of spectroscopic signals (mathematically, tensors) plays an important role on the theory of nonlinear spectroscopy. The rotational averaging of Cartesian tensors has been extensively considered last half-century\cite{} specially for tensors of rank lower than 9. However, complication of the problem associated with rotational averaging becomes more serious as rank of tensors increases. In order to overcome such
difficulty arisen from high-rank tensor averaging, we make our physical model as simple as possible.
 
One of the most simplest experimental realization of CARS in optically active medium would be the experimental configuration where linearly polarized three co-linear input pulse interact with the sample generating anti-Stokes light in the forward direction $\mathbf{e}_z$, and measurable quantity is the circularly polarized components of scattered anti-Stokes light. The reason for that choice is to avoid an accumulated error due to quarter wave plates to create circularly polarized incident lights. The optimal way is to select only one of the four beams to be circularly polarized (for our case anti-Stokes beam) in order to be free from error as much as possible.
For further simplification, we ignore dispersion of the sample and assume all four wave vectors are co-linear $\mathbf{k}_1=\mathbf{k}_2=\mathbf{k}_3=\mathbf{k}_4=\mathbf{e}_z$.

When three input laser fields have vertical polarization $\mathbf{e}_1=\mathbf{e}_2=\mathbf{e}_3=\mathbf{e}_x$ and right and left circularly polarized components of scattered anti-Stokes field are of interest, we denote this experimental configuration as VVVR and VVVL. Here, V stands for vertical polarization $\mathbf{e}_x$. We adopt a convention $\mathbf{e}_R=\mathbf{e}_x-i\mathbf{e}_y$ and $\mathbf{e}_L=\mathbf{e}_x+i\mathbf{e}_y$ for right and left circular polarization unit vectors.

Norm of the scattering matrix $M_{FI}$ can be found from Eq.~\ref{general} as follows
\begin{align}\label{general_vvvr}
&\vert M_{FI}\vert^2
=\pi^2\rho_s^2
\left(
\frac{\hbar c}{2\varepsilon_0 V}
\right)^4
k_1k_2k_3k_4 n_1n_3(n_2+1)(n_4+1)\nonumber\\
&\left(
\frac{1}{2}
\alpha_{xx}(-\omega_3,\omega_4)
\alpha_{xx}(-\omega_1,\omega_2)
\alpha_{xx}(-\omega_3,\omega_4)
\alpha_{xx}(-\omega_1,\omega_2)\right.
\nonumber\\
&+
\frac{1}{2}
\alpha_{yx}(-\omega_3,\omega_4)
\alpha_{xx}(-\omega_1,\omega_2)
\alpha_{yx}(-\omega_3,\omega_4)
\alpha_{xx}(-\omega_1,\omega_2)\nonumber\\
&+
\frac{1}{c}
G_{yy}'(-\omega_3,\omega_4)
\alpha_{xx}(-\omega_1,\omega_2)
\alpha_{xx}(-\omega_3,\omega_4)
\alpha_{xx}(-\omega_1,\omega_2)\nonumber\\
&+
\frac{1}{c}
G_{xx}'(-\omega_3,\omega_4)
\alpha_{xx}(-\omega_1,\omega_2)
\alpha_{xx}(-\omega_3,\omega_4)
\alpha_{xx}(-\omega_1,\omega_2)\nonumber\\
&
-\frac{k_3}{3}
A_{y,xz}(-\omega_3,\omega_4)
\alpha_{xx}(-\omega_1,\omega_2)
\alpha_{xx}(-\omega_3,\omega_4)
\alpha_{xx}(-\omega_1,\omega_2)\nonumber\\
&+\frac{k_3}{3}
A_{x,xz}(-\omega_3,\omega_4)
\alpha_{xx}(-\omega_1,\omega_2)
\alpha_{yx}(-\omega_3,\omega_4)
\alpha_{xx}(-\omega_1,\omega_2)\nonumber\\
&+
\frac{k_4}{3}
A_{x,yz}(-\omega_3,\omega_4)
\alpha_{xx}(-\omega_1,\omega_2)
\alpha_{xx}(-\omega_3,\omega_4)
\alpha_{xx}(-\omega_1,\omega_2)\nonumber\\
&\left.-\frac{k_4}{3}
A_{x,xz}(-\omega_3,\omega_4)
\alpha_{xx}(-\omega_1,\omega_2)
\alpha_{yx}(-\omega_3,\omega_4)
\alpha_{xx}(-\omega_1,\omega_2).
\right)
\end{align}
Interestingly, during the calculation of Eq.~\ref{general_vvvr} we observed that optical activity terms are only due to probe and anti-Stokes transitions i.e. pump and Stokes fields provide only electric dipole contribution. Furthermore, we also observed that this behavior is also present for VVHR and VVHL configuration where the letter H stands for horizontal polarization unit vector $\mathbf{e}_y$ for probe field $\mathbf{k}_3$.

We interested in measurable quantity $\Delta$ defined below in the same manner as circular intensity difference in the theory of Raman optical activity:
\begin{align}
\Delta=(T^{(R)}_{CARS}-T^{(L)}_{CARS})/(T^{(R)}_{CARS}+T^{(L)}_{CARS})
\end{align}
where $T^{(R)}_{CARS}$ is a transition rate for VVVR measurement while $T^{(L)}_{CARS}$ for VVVL. The obtained result for $\Delta$ is given by
\begin{widetext}
\begin{align}
\Delta
&=
\left[\frac{1}{c}
\left(
\frac{514}{5145}g_0^{(11)}
-\frac{2056}{15435}g_0^{(12)}
-\frac{341}{5145}g_0^{(21)}
+\frac{682}{15435}g_0^{(22)}
+\frac{16}{525}g_2^{(11)}
-\frac{8}{105}g_2^{(12)}\right.\right.\nonumber\\
&-\left.\frac{1}{105}g_2^{(21)}
+\frac{4}{315}g_2^{(22)}
-\frac{1}{105}g_2^{(31)}
+\frac{4}{315}g_2^{(32)}
+\frac{2}{525}g_2^{(41)}
-\frac{1}{105}g_2^{(42)}
+\frac{1}{315}g_4^{(11)}
\right)\nonumber\\
&+
\frac{1}{3c}\left(
\frac{7853}{92610}k_{0,\omega_3}^{(21)}
-\frac{7853}{138915}k_{0,\omega_3}^{(22)}
+\frac{5}{378}k_{2,\omega_3}^{(21)}
-\frac{10}{567}k_{2,\omega_3}^{(22)}
+\frac{1}{105}k_{2,\omega_3}^{(31)}
-\frac{4}{315}k_{2,\omega_3}^{(32)}\right.\nonumber\\
&\left.-\frac{2}{525}k_{2,\omega_3}^{(41)}
+\frac{1}{105}k_{2,\omega_3}^{(42)}
-\frac{1}{315}k_{4,\omega_3}^{(11)}
\right)
-\frac{1}{3c}
\left.\left(
\frac{1}{54}k_{0,\omega_4}^{(21)}
-\frac{1}{81}k_{0,\omega_4}^{(22)}
+\frac{1}{270}k_{2,\omega_4}^{(21)}
-\frac{2}{405}k_{2,\omega_4}^{(22)}
\right)\right]\nonumber\\
&\left.\middle/
\left(\frac{1}{120}a_0^{(11)}
-\frac{1}{30}a_0^{(12)}
-\frac{7}{60}a_0^{(21)}
+\frac{7}{90}a_0^{(22)}
+\frac{1}{525}a_2^{(11)}
-\frac{1}{210}a_2^{(12)}
-\frac{11}{840}a_2^{(21)}
+\frac{11}{630}a_2^{(22)}
+\frac{2}{315}a_4^{(11)}\right)\right.
\end{align}
where $a$'s, $g$'s and $k$'s are natural invariants of various type of products of $\alpha_{ij}$, $G'_{ij}$ and $A_{i,jn}$ tensors, respectively. Explicit forms and interpretation of these natural invariants are given in Appendix~\ref{app_b}.
In the approximation of $\omega_3\approx \omega_4$, 
\begin{align}\label{signal}
\Delta
&=
\left[\frac{1}{c}
\left(
\frac{514}{5145}g_0^{(11)}
-\frac{2056}{15435}g_0^{(12)}
-\frac{341}{5145}\left(g_0^{(21)}-k_0^{(21)}/3\right)
+\frac{682}{15435}\left(g_0^{(22)}-k_0^{(22)}/3\right)
+\frac{16}{525}g_2^{(11)}\right.\right.\nonumber\\
&-\frac{8}{105}g_2^{(12)}
-\frac{1}{105}\left(g_2^{(21)}-k_2^{(21)}/3\right)
+\frac{4}{315}\left(g_2^{(22)}-k_2^{(22)}/3\right)
-\frac{1}{105}\left(g_2^{(31)}-k_2^{(31)}/3\right)\nonumber\\
&+\left.\frac{4}{315}\left(g_2^{(32)}-k_2^{(32)}/3\right)
+\frac{2}{525}\left(g_2^{(41)}-k_2^{(41)}/3\right)
-\frac{1}{105}\left(g_2^{(42)}-k_2^{(42)}/3\right)
+\frac{1}{315}\left(g_4^{(11)}-k_4^{(11)}/3\right)
\right)
\left.\right]\nonumber\\
&\left.\middle/
\left(\frac{1}{120}a_0^{(11)}
-\frac{1}{30}a_0^{(12)}
-\frac{7}{60}a_0^{(21)}
+\frac{7}{90}a_0^{(22)}
+\frac{1}{525}a_2^{(11)}
-\frac{1}{210}a_2^{(12)}
-\frac{11}{840}a_2^{(21)}
+\frac{11}{630}a_2^{(22)}
+\frac{2}{315}a_4^{(11)}\right)\right.
\end{align}
\end{widetext}
Here, natural invariants $k_0^{(11)}$, $k_0^{(12)}$, $k_2^{(11)}$ and $k_2^{(12)}$ vanish as we show in Appendix~\ref{app_b}. It is clear to see that all prefactors for natural invariants $g$ and $k$ in Eq.~\ref{signal} are the same even though they are calculated using rotational average of different rank (eighth- and ninth-rank) tensors. This implies the correctness of our rigorous calculations.

\section{Conclusions and Discussions}
In summary, we develop quantum electrodynamic theory of CARS from randomly oriented chiral molecules. The explicit expressions of post-selected chiral signals for VVVR and VVVL polarization configuration in terms of natural invariants of corresponding tensor products are found. The obtained expression would be extremely helpful for comparing theoretically predicted and experimentally obtained CARS spectra for chiral discrimination, ones polarizability and optical activity tensors are found by first principle calculation for specific molecule. It is worth to mention here that we model our theory as simple as possible and just select one simplest polarization configuration VVVR and VVVL.

Our result has two crucial constraints listed below:
\begin{enumerate}
\item Phase matching condition for wave vectors. Due to this condition there must be non-zero angle between wave vectors $\mathbf{k}_i$ of incident and scattered beams in dispersive medium. The angles between wave vectors typically vary from $1^\text{o}$ to $3^\text{o}$ in most samples in the gas and liquid phases\cite{Bjarnason1980}. Therefore, our main result of this work should be slightly modified because of the phase matching condition in dispersive sample. 
\item Constraint due to resonance condition. As we wrote in the main text of this paper, our result only valid for off-resonance condition otherwise complex nature of polarizability and optical activity tensors is needed to be considered.
\end{enumerate}

Despite above mentioned two constraints, our expression for $\Delta$ can be applicable for computations of CARS spectra in optically active medium. We hope that future works will extend this results and makes it free from the constraints mentioned above.

\begin{acknowledgments}
We are grateful to the Air Force Office of Scientific Research (Award No. FA9550-18-1-0141), the Office of Naval Research (Award No. N00014-16-1-3054), and the Robert A. Welch Foundation (Grant No. A-1261).
\end{acknowledgments}

\appendix
\section{Optical activity tensors}\label{app_a}
Optical activity tensors are given by
\begin{align}
G_{kl}^{(1)}(-\omega_1,\omega_2)
&=
\sum_{t}
\left\{
\frac{\langle s\vert\hat{\mu}_k\vert t\rangle
\langle t\vert\hat{m}_l\vert g\rangle}{[E_{tg}-\hbar \omega_1-i\epsilon]}
+
\frac{\langle s\vert\hat{m}_l\vert t\rangle
\langle t\vert\hat{\mu}_k\vert g\rangle}{[E_{tg}+\hbar \omega_2+i\epsilon]}
\right\},\\
G_{kl}^{(2)}(-\omega_1,\omega_2)
&=
\sum_{t}
\left\{
\frac{\langle s\vert\hat{m}_k\vert t\rangle
\langle t\vert\hat{\mu}_l\vert g\rangle}{[E_{tg}-\hbar \omega_1-i\epsilon]}
+
\frac{\langle s\vert\hat{\mu}_l\vert t\rangle
\langle t\vert\hat{m}_k\vert g\rangle}{[E_{tg}+\hbar \omega_2+i\epsilon]}
\right\},\\
G_{ij}^{(1)}(-\omega_3,\omega_4)
&=
\sum_{r}
\left\{
\frac{
\langle f\vert\hat{\mu}_i\vert r\rangle
\langle r\vert\hat{m}_j\vert s\rangle
}{[E_{rs} - \hbar\omega_3-i\epsilon]}
+
\frac{
\langle f\vert\hat{m}_j\vert r\rangle
\langle r\vert\hat{\mu}_i\vert s\rangle
}{[E_{rs} + \hbar\omega_4+i\epsilon]}
\right\},\\
G_{ij}^{(2)}(-\omega_3,\omega_4)
&=
\sum_{r}
\left\{
\frac{
\langle f\vert\hat{m}_i\vert r\rangle
\langle r\vert\hat{\mu}_j\vert s\rangle
}{[E_{rs} - \hbar\omega_3-i\epsilon]}
+
\frac{
\langle f\vert\hat{\mu}_j\vert r\rangle
\langle r\vert\hat{m}_i\vert s\rangle
}{[E_{rs} + \hbar\omega_4+i\epsilon]}
\right\},
\end{align}
and
\begin{align}
A^{(1)}_{k,ln}(-\omega_1,\omega_2)
&=
\sum_{t}
\left\{
\frac{\langle s\vert\hat{\mu}_k\vert t\rangle
\langle t\vert\hat{q}_{ln}\vert g\rangle}{[E_{tg}-\hbar \omega_1-i\epsilon]}
+
\frac{\langle s\vert\hat{q}_{ln}\vert t\rangle
\langle t\vert\hat{\mu}_k\vert g\rangle}{[E_{tg}+\hbar \omega_2+i\epsilon]}
\right\},\\
A^{(2)}_{l,kn}(-\omega_1,\omega_2)
&=
\sum_{t}
\left\{
\frac{\langle s\vert\hat{q}_{kn}\vert t\rangle
\langle t\vert\hat{\mu}_{l}\vert g\rangle}{[E_{tg}-\hbar \omega_1-i\epsilon]}
+
\frac{\langle s\vert\hat{\mu}_{l}\vert t\rangle
\langle t\vert\hat{q}_{kn}\vert g\rangle}{[E_{tg}+\hbar \omega_2+i\epsilon]}
\right\},\\
A^{(1)}_{i,jn}(-\omega_3,\omega_4)
&=
\sum_{r}
\left\{
\frac{
\langle f\vert\hat{\mu}_i\vert r\rangle
\langle r\vert\hat{q}_{jn}\vert s\rangle
}{[E_{rs} - \hbar\omega_3-i\epsilon]}
+
\frac{
\langle f\vert\hat{q}_{jn}\vert r\rangle
\langle r\vert\hat{\mu}_i\vert s\rangle
}{[E_{rs} + \hbar\omega_4+i\epsilon]}
\right\},\\
A^{(2)}_{j,in}(-\omega_3,\omega_4)
&=
\sum_{r}
\left\{
\frac{
\langle f\vert\hat{q}_{in}\vert r\rangle
\langle r\vert\hat{\mu}_j\vert s\rangle
}{[E_{rs} - \hbar\omega_3-i\epsilon]}
+
\frac{
\langle f\vert\hat{\mu}_j\vert r\rangle
\langle r\vert\hat{q}_{in}\vert s\rangle
}{[E_{rs} + \hbar\omega_4+i\epsilon]}
\right\}.
\end{align}
Above tensors are found by replacing one of the electric dipole operator in Eqs.~\ref{poltensor1},\ref{poltensor2} by either magnetic dipole or electric quadrupole operators.

\section{Details of isotropic average}\label{app_b}
In this appendix, we presented details of calculation for VVVR configuration. Its counterpart--VVVL configuration provides the same term for electric dipole contribution whereas the equal terms in magnitude with opposite signs for optical activity contributions. Therefore, we  do not show details for VVVL configuration.
\subsection{Isotropic average of electric dipole term}\label{app_electric_dipole}
The first two terms in Eq.~\ref{general_vvvr} associated to electric dipole transition are components of eight-rank tensor. Using overcomplete set of isotropic basis tensors\cite{Andrews1981} of rank 8 we obtain rotational average of these terms as follows 
\begin{align}\label{electricdipole1}
&\frac{1}{2}
\langle
\alpha_{xx}(-\omega_3,\omega_4)
\alpha_{xx}(-\omega_1,\omega_2)
\alpha_{xx}(-\omega_3,\omega_4)
\alpha_{xx}(-\omega_1,\omega_2)
\rangle
\nonumber\\
&+
\frac{1}{2}
\langle
\alpha_{yx}(-\omega_3,\omega_4)
\alpha_{xx}(-\omega_1,\omega_2)
\alpha_{yx}(-\omega_3,\omega_4)
\alpha_{xx}(-\omega_1,\omega_2)
\rangle\nonumber\\
&=\frac{1}{3780}
\left(
[\alpha]_1
+8[\alpha]_2
+16[\alpha]_3
+2[\alpha]_4
+8[\alpha]_5
+52[\alpha]_6\right.\nonumber\\
&\left.+104[\alpha]_7
+16[\alpha]_8
+11[\alpha]_9
+22[\alpha]_{10}
\right),
\end{align}
where isotropic invariants are given by
\begin{align}\label{invariantselectricdipole}
[\alpha ]_1&=
\alpha_{ii}(-\omega_3,\omega_4)
\alpha_{jj}(-\omega_1,\omega_2)
\alpha_{kk}(-\omega_3,\omega_4)
\alpha_{ll}(-\omega_1,\omega_2),\nonumber\\
[\alpha ]_2&=
\alpha_{ii}(-\omega_3,\omega_4)
\alpha_{jj}(-\omega_1,\omega_2)
\alpha_{kl}(-\omega_3,\omega_4)
\alpha_{kl}(-\omega_1,\omega_2),\nonumber\\
[\alpha ]_3&=
\alpha_{ii}(-\omega_3,\omega_4)
\alpha_{jk}(-\omega_1,\omega_2)
\alpha_{jl}(-\omega_3,\omega_4)
\alpha_{kl}(-\omega_1,\omega_2),\nonumber\\
[\alpha ]_4&=
\alpha_{ii}(-\omega_3,\omega_4)
\alpha_{jk}(-\omega_1,\omega_2)
\alpha_{ll}(-\omega_3,\omega_4)
\alpha_{jk}(-\omega_1,\omega_2),\nonumber\\
[\alpha ]_5&=
\alpha_{ij}(-\omega_3,\omega_4)
\alpha_{ij}(-\omega_1,\omega_2)
\alpha_{kl}(-\omega_3,\omega_4)
\alpha_{kl}(-\omega_1,\omega_2),\nonumber\\
[\alpha ]_6&=
\alpha_{ij}(-\omega_3,\omega_4)
\alpha_{ik}(-\omega_1,\omega_2)
\alpha_{jk}(-\omega_3,\omega_4)
\alpha_{ll}(-\omega_1,\omega_2),\nonumber\\
[\alpha ]_7&=
\alpha_{ij}(-\omega_3,\omega_4)
\alpha_{ik}(-\omega_1,\omega_2)
\alpha_{jl}(-\omega_3,\omega_4)
\alpha_{kl}(-\omega_1,\omega_2),\nonumber\\
[\alpha ]_8&=
\alpha_{ij}(-\omega_3,\omega_4)
\alpha_{ik}(-\omega_1,\omega_2)
\alpha_{kl}(-\omega_3,\omega_4)
\alpha_{jl}(-\omega_1,\omega_2),\nonumber\\
[\alpha ]_9&=
\alpha_{ij}(-\omega_3,\omega_4)
\alpha_{kk}(-\omega_1,\omega_2)
\alpha_{ij}(-\omega_3,\omega_4)
\alpha_{ll}(-\omega_1,\omega_2),\nonumber\\
[\alpha ]_{10}&=
\alpha_{ij}(-\omega_3,\omega_4)
\alpha_{kl}(-\omega_1,\omega_2)
\alpha_{ij}(-\omega_3,\omega_4)
\alpha_{kl}(-\omega_1,\omega_2).
\end{align}
Note that the symmetric property of the tensor $\alpha_{ij}$ under permutation of two its indices is used in the calculation of Eq.~\ref{electricdipole1} and it actually reduces the number of distinct isotropic invariants from 105 to 10.

However, ten invariants given by Eq.~\ref{invariantselectricdipole} are not all independent since we use overcomplete isotropic basis set. There are 14 Young tableau of the shape $(2,2,2,2)$, and checking one by one we found that the standard Young tableau given in Fig.~\ref{youngtableaux}
\begin{figure}
\begin{ytableau}
1 & 2\\
3 & 4\\
5 & 6\\
7 & 8\\
\end{ytableau}
\caption{Standard Young tableau that provides the linear dependence between isotropic invariants $[\alpha]_i$, $[G']_i$ and $[A]_i$.}\label{youngtableaux}
\end{figure}
produces the desired relation between the invariants $[\alpha]_i$. Other 13 standard Young tableaux yield trivial relation $0=0$. The obtained result is written as
\begin{align}\label{relationelectricdipole}
&[\alpha]_1
-4[\alpha]_2
+4[\alpha]_3
-[\alpha]_4
+2[\alpha]_5
+4[\alpha]_6\nonumber\\
&-4[\alpha]_7
-2[\alpha]_8
-[\alpha]_9
+[\alpha]_{10}
=
0.
\end{align}
The relation Eq.~\ref{relationelectricdipole} allow us to express one of the isotropic tensors in terms of others, and as a result we have 9 independent isotropic invariants. These 9 isotropic invariants can be chosen arbitrarily from full 10 set since there in no physically meaningful difference between ten isotropic invariants $[\alpha]_i$. 

However, it is more acceptable to express rotational averages in terms of natural invariants \cite{Jerphagnon1978,Ford2018} that are more physically meaningful and associated to symmetry types of high-rank tensors composed of polarizability and optical activity tensors. To do so, let us introduce new fourth-rank tensors
\begin{align}\label{tensorstt}
\bar{T}_{[(ij)(kl)]}&=\alpha_{ij}(-\omega_3,\omega_4)\alpha_{kl}(-\omega_3,\omega_4),\nonumber\\
T_{[(mo)(pq)]}&=\alpha_{mo}(-\omega_1,\omega_2)\alpha_{pq}(-\omega_1,\omega_2).
\end{align}
These tensors are symmetric under permutation of its first two indices, its last two indices and these pairs of indices. Spectral decomposition of fourth-rank Cartesian tensors into its irreducible subspaces is well-known, and it is explicitly found by D.~L.~Andrews et al\cite{Andrews1982}. According to their result, there are only 5 terms for the tensors given by Eq.~\ref{tensorstt} as follows
\begin{align}
\bar{T}_{[(ij)(kl)]}=\bar{T}^{(0,1)}_{[(ij)(kl)]}+\bar{T}^{(0,2)}_{[(ij)(kl)]}+\bar{T}^{(2,1)}_{[(ij)(kl)]}+\bar{T}^{(2,2)}_{[(ij)(kl)]}+\bar{T}^{(4,1)}_{[(ij)(kl)]},
\end{align}
where the first number in superscript refers to weight and the second refers to a seniority. Since the tensors of different weights are orthogonal to each other, contraction of tensors $\bar{T}_{[(ij)(kl)]}$ and $T_{[(mo)(pq)]}$ yields
\begin{align}\label{contracted1}
&\bar{T}_{[(ij)(kl)]}T_{[(ij)(kl)]}\nonumber\\
&=\bar{T}^{(0,1)}_{[(ij)(kl)]}T^{(0,1)}_{[(ij)(kl)]}+\bar{T}^{(0,1)}_{[(ij)(kl)]}T^{(0,2)}_{[(ij)(kl)]}\nonumber\\
&+\bar{T}^{(0,2)}_{[(ij)(kl)]}T^{(0,1)}_{[(ij)(kl)]}+\bar{T}^{(0,2)}_{[(ij)(kl)]}T^{(0,2)}_{[(ij)(kl)]}\nonumber\\
&+\bar{T}^{(2,1)}_{[(ij)(kl)]}T^{(2,1)}_{[(ij)(kl)]}+\bar{T}^{(2,1)}_{[(ij)(kl)]}T^{(2,2)}_{[(ij)(kl)]}\nonumber\\
&+\bar{T}^{(2,2)}_{[(ij)(kl)]}T^{(2,1)}_{[(ij)(kl)]}+\bar{T}^{(2,2)}_{[(ij)(kl)]}T^{(2,2)}_{[(ij)(kl)]}\nonumber\\
&+\bar{T}^{(4,1)}_{[(ij)(kl)]}T^{(4,1)}_{[(ij)(kl)]}.
\end{align}
Each term in Eq.~\ref{contracted1} represents natural invariant of given weight and seniority, and
we introduce a notation $a_J^{(\tau_1\tau_2)}=\bar{T}^{(J,\tau_1)}_{[(ij)(kl)]}T^{(J,\tau_2)}_{[(ij)(kl)]}$ for these natural invariants. Then Eq.~\ref{electricdipole1} has following form in terms of natural invariants:
\begin{align}
&\frac{1}{2}
\langle
\alpha_{xx}(-\omega_3,\omega_4)
\alpha_{xx}(-\omega_1,\omega_2)
\alpha_{xx}(-\omega_3,\omega_4)
\alpha_{xx}(-\omega_1,\omega_2)
\rangle
\nonumber\\
&+
\frac{1}{2}
\langle
\alpha_{yx}(-\omega_3,\omega_4)
\alpha_{xx}(-\omega_1,\omega_2)
\alpha_{yx}(-\omega_3,\omega_4)
\alpha_{xx}(-\omega_1,\omega_2)
\rangle\nonumber\\
&=
\frac{1}{120}a_0^{(11)}
-\frac{1}{30}a_0^{(12)}
-\frac{7}{60}a_0^{(21)}
+\frac{7}{90}a_0^{(22)}
+\frac{1}{525}a_2^{(11)}\nonumber\\
&-\frac{1}{210}a_2^{(12)}
-\frac{11}{840}a_2^{(21)}
+\frac{11}{630}a_2^{(22)}
+\frac{2}{315}a_4^{(11)}
\end{align}
where
\begin{align}
a_0^{(11)}&=\frac{2}{15}
[\alpha]_1,\nonumber\\
a_0^{(12)}&=-\frac{1}{15}
[\alpha]_4,\nonumber\\
a_0^{(22)}&=\frac{1}{5}
[\alpha]_{10},\nonumber\\
a_0^{(21)}&=
-\frac{1}{15}
[\alpha]_9,\nonumber\\
a_2^{(11)}&=
-\frac{10}{21}
[\alpha]_1+
\frac{10}{7}
[\alpha]_2,\nonumber\\
a_2^{(22)}&=
\frac{12}{7}
[\alpha]_7
-
\frac{4}{7}
[\alpha]_{10},\nonumber\\
a_2^{(12)}&=
-\frac{8}{7}
[\alpha]_3
+
\frac{8}{21}
[\alpha]_4,\nonumber\\
a_2^{(21)}&=
-\frac{8}{7}
[\alpha]_6
+
\frac{8}{21}
[\alpha]_9,\nonumber\\
a_4^{(11)}&=
-\frac{11}{70}[\alpha]_1
+\frac{4}{7}[\alpha]_2
-\frac{6}{7}[\alpha]_3
+\frac{13}{70}[\alpha]_4
-\frac{6}{7}[\alpha]_6\nonumber\\
&+\frac{2}{7}[\alpha]_7
+[\alpha]_8
+\frac{13}{70}[\alpha]_9
-\frac{9}{70}[\alpha]_{10}.
\end{align}

\subsection{Isotropic average of magnetic dipole term}
The third and fourth terms in Eq.~\ref{general_vvvr} refer to magnetic dipole interaction. The main procedure of calculation is the same as what we did in appendix~\ref{app_electric_dipole}. However, the only difference here is that the tensor $G'_{ij}$ is not a symmetric tensor unlike $\alpha_{ij}$. For this reason, we have more, actually, 14 isotropic invariants as follows:
\begin{align}
[G']_1&=
G_{ii}'(-\omega_3,\omega_4)
\alpha_{jj}(-\omega_1,\omega_2)
\alpha_{kk}(-\omega_3,\omega_4)
\alpha_{ll}(-\omega_1,\omega_2),\nonumber\\
[G']_2&=
G_{ii}'(-\omega_3,\omega_4)
\alpha_{jj}(-\omega_1,\omega_2)
\alpha_{kl}(-\omega_3,\omega_4)
\alpha_{kl}(-\omega_1,\omega_2),\nonumber\\
[G']_3&=
G_{ii}'(-\omega_3,\omega_4)
\alpha_{jk}(-\omega_1,\omega_2)
\alpha_{jl}(-\omega_3,\omega_4)
\alpha_{kl}(-\omega_1,\omega_2),\nonumber\\
[G']_4&=
G_{ii}'(-\omega_3,\omega_4)
\alpha_{jk}(-\omega_1,\omega_2)
\alpha_{ll}(-\omega_3,\omega_4)
\alpha_{jk}(-\omega_1,\omega_2),\nonumber\\
[G']_5&=
G_{ij}'(-\omega_3,\omega_4)
\alpha_{ij}(-\omega_1,\omega_2)
\alpha_{kk}(-\omega_3,\omega_4)
\alpha_{ll}(-\omega_1,\omega_2),\nonumber\\
[G']_6&=
G_{ij}'(-\omega_3,\omega_4)
\alpha_{ij}(-\omega_1,\omega_2)
\alpha_{kl}(-\omega_3,\omega_4)
\alpha_{kl}(-\omega_1,\omega_2),\nonumber\\
[G']_7&=
G_{ij}'(-\omega_3,\omega_4)
\alpha_{ik}(-\omega_1,\omega_2)
\alpha_{jk}(-\omega_3,\omega_4)
\alpha_{ll}(-\omega_1,\omega_2),\nonumber\\
[G']_8&=
G_{ij}'(-\omega_3,\omega_4)
\alpha_{ik}(-\omega_1,\omega_2)
\alpha_{jl}(-\omega_3,\omega_4)
\alpha_{kl}(-\omega_1,\omega_2),\nonumber\\
[G']_9&=
G_{ij}'(-\omega_3,\omega_4)
\alpha_{ik}(-\omega_1,\omega_2)
\alpha_{kl}(-\omega_3,\omega_4)
\alpha_{jl}(-\omega_1,\omega_2),\nonumber\\
[G']_{10}&=
G_{ij}'(-\omega_3,\omega_4)
\alpha_{ik}(-\omega_1,\omega_2)
\alpha_{ll}(-\omega_3,\omega_4)
\alpha_{jk}(-\omega_1,\omega_2),\nonumber\\
[G']_{11}&=
G_{ij}'(-\omega_3,\omega_4)
\alpha_{jk}(-\omega_1,\omega_2)
\alpha_{ik}(-\omega_3,\omega_4)
\alpha_{ll}(-\omega_1,\omega_2),\nonumber\\
[G']_{12}&=
G_{ij}'(-\omega_3,\omega_4)
\alpha_{jk}(-\omega_1,\omega_2)
\alpha_{il}(-\omega_3,\omega_4)
\alpha_{kl}(-\omega_1,\omega_2),\nonumber\\
[G']_{13}&=
G_{ij}'(-\omega_3,\omega_4)
\alpha_{kk}(-\omega_1,\omega_2)
\alpha_{ij}(-\omega_3,\omega_4)
\alpha_{ll}(-\omega_1,\omega_2),\nonumber\\
[G']_{14}&=
G_{ij}'(-\omega_3,\omega_4)
\alpha_{kl}(-\omega_1,\omega_2)
\alpha_{ij}(-\omega_3,\omega_4)
\alpha_{kl}(-\omega_1,\omega_2).
\end{align}
These 14 invariants form overcomplete set of isotropic invariants. Magnetic dipole contribution averaged over three-dimensional rotation is obtained in terms of isotropic invariants $[G']_i$ as follows:
\begin{align}\label{magneticdipole1}
&\frac{1}{c}
\langle G_{yy}'(-\omega_3,\omega_4)
\alpha_{xx}(-\omega_1,\omega_2)
\alpha_{xx}(-\omega_3,\omega_4)
\alpha_{xx}(-\omega_1,\omega_2)\rangle\nonumber\\
&+
\frac{1}{c}
\langle G_{xx}'(-\omega_3,\omega_4)
\alpha_{xx}(-\omega_1,\omega_2)
\alpha_{xx}(-\omega_3,\omega_4)
\alpha_{xx}(-\omega_1,\omega_2)\rangle\nonumber\\
&=
\frac{1}{c}
\left(
\frac{40}{7560}[G']_1
+\frac{160}{7560}[G']_2
+\frac{320}{7560}[G']_3
+\frac{80}{7560}[G']_4\right.\nonumber\\
&+\frac{16}{7560}[G']_5
+\frac{32}{7560}[G']_6
+\frac{32}{7560}[G']_7
+\frac{64}{7560}[G']_8\nonumber\\
&+\frac{64}{7560}[G']_9
+\frac{32}{7560}[G']_{10}
+\frac{32}{7560}[G']_{11}
+\frac{64}{7560}[G']_{12}\nonumber\\
&+\left.\frac{8}{7560}[G']_{13}
+\frac{16}{7560}[G']_{14}
\right).
\end{align}
Similarly as electric dipole contribution part, 
the Young tableau shown in Fig.~\ref{youngtableaux} yields a relation
\begin{align}\label{relationmagneticdipole}
&[G']_1
-2[G']_2
+2[G']_3
-[G']_4
-2[G']_5
+2[G']_6
+2[G']_7\nonumber\\
&-2[G']_8
-2[G']_9
+2[G']_{10}
+2[G']_{11}
-2[G']_{12}
-[G']_{13}\nonumber\\
&+[G']_{14}
=0.
\end{align}
For natural invariants, the tensor $T_{[(mo)(pq)]}$ still have its form defined by Eq.~\ref{tensorstt} while $\bar{T}$ is redefined as
\begin{align}
\bar{T}_{[ij(kl)]}&=G'_{ij}(-\omega_3,\omega_4)\alpha_{kl}(-\omega_3,\omega_4),
\end{align}
and it is symmetric under permutation of only last two indices. Spectral decomposition of the tensor $\bar{T}_{[ij(kl)]}$ is given by
\begin{align}
&\bar{T}_{[ij(kl)]}\nonumber\\
&=
\bar{T}^{(0,1)}_{[ij(kl)]}
+\bar{T}^{(0,2)}_{[ij(kl)]}
+\bar{T}^{(1,1)}_{[ij(kl)]}
+\bar{T}^{(1,2)}_{[ij(kl)]}\nonumber\\
&+\bar{T}^{(1,3)}_{[ij(kl)]}
+\bar{T}^{(2,1)}_{[ij(kl)]}
+\bar{T}^{(2,2)}_{[ij(kl)]}
+\bar{T}^{(2,3)}_{[ij(kl)]}\nonumber\\
&+\bar{T}^{(2,4)}_{[ij(kl)]}
+\bar{T}^{(3,1)}_{[ij(kl)]}
+\bar{T}^{(3,2)}_{[ij(kl)]}
+\bar{T}^{(4,1)}_{[ij(kl)]}.
\end{align}
Contraction of the tensors $\bar{T}_{[ij(kl)]}$ and $T_{[(ij)(kl)]}$ are
\begin{align}
&\bar{T}_{[(ij)(kl)]}T_{[(ij)(kl)]}\nonumber\\
&=\bar{T}^{(0,1)}_{[(ij)(kl)]}T^{(0,1)}_{[(ij)(kl)]}+\bar{T}^{(0,1)}_{[(ij)(kl)]}T^{(0,2)}_{[(ij)(kl)]}\nonumber\\
&+\bar{T}^{(0,2)}_{[(ij)(kl)]}T^{(0,1)}_{[(ij)(kl)]}+\bar{T}^{(0,2)}_{[(ij)(kl)]}T^{(0,2)}_{[(ij)(kl)]}\nonumber\\
&+\bar{T}^{(2,1)}_{[(ij)(kl)]}T^{(2,1)}_{[(ij)(kl)]}+\bar{T}^{(2,1)}_{[(ij)(kl)]}T^{(2,2)}_{[(ij)(kl)]}\nonumber\\
&+\bar{T}^{(2,2)}_{[(ij)(kl)]}T^{(2,1)}_{[(ij)(kl)]}+\bar{T}^{(2,2)}_{[(ij)(kl)]}T^{(2,2)}_{[(ij)(kl)]}\nonumber\\
&+\bar{T}^{(2,3)}_{[(ij)(kl)]}T^{(2,1)}_{[(ij)(kl)]}+\bar{T}^{(2,3)}_{[(ij)(kl)]}T^{(2,2)}_{[(ij)(kl)]}\nonumber\\
&+\bar{T}^{(2,4)}_{[(ij)(kl)]}T^{(2,1)}_{[(ij)(kl)]}+\bar{T}^{(2,4)}_{[(ij)(kl)]}T^{(2,2)}_{[(ij)(kl)]}\nonumber\\
&+\bar{T}^{(4,1)}_{[(ij)(kl)]}T^{(4,1)}_{[(ij)(kl)]}.
\end{align}
With the short notation as $g_J^{(\tau_1\tau_2)}=\bar{T}^{(J,\tau_1)}_{[(ij)(kl)]}T^{(J,\tau_2)}_{[(ij)(kl)]}$, the magnetic dipole contribution Eq.~\ref{magneticdipole1} becomes as follows:
\begin{align}
&\frac{1}{c}
\langle
G_{yy}'(-\omega_3,\omega_4)
\alpha_{xx}(-\omega_1,\omega_2)
\alpha_{xx}(-\omega_3,\omega_4)
\alpha_{xx}(-\omega_1,\omega_2)\rangle\nonumber\\
&+
\frac{1}{c}
\langle
G_{xx}'(-\omega_3,\omega_4)
\alpha_{xx}(-\omega_1,\omega_2)
\alpha_{xx}(-\omega_3,\omega_4)
\alpha_{xx}(-\omega_1,\omega_2)\rangle\nonumber\\
&=
\frac{1}{c}
\left(
\frac{514}{5145}g_0^{(11)}
-\frac{2056}{15435}g_0^{(12)}
-\frac{341}{5145}g_0^{(21)}
+\frac{682}{15435}g_0^{(22)}\right.\nonumber\\
&+\frac{16}{525}g_2^{(11)}
-\frac{8}{105}g_2^{(12)}
-\frac{1}{105}g_2^{(21)}
+\frac{4}{315}g_2^{(22)}
-\frac{1}{105}g_2^{(31)}\nonumber\\
&+\left.\frac{4}{315}g_2^{(32)}
+\frac{2}{525}g_2^{(41)}
-\frac{1}{105}g_2^{(42)}
+\frac{1}{315}g_4^{(11)}
\right),
\end{align}
where
\begin{align}
g_0^{(11)}=&\frac{2}{15}
[G']_1,\nonumber\\
g_0^{(12)}=&-\frac{1}{15}
[G']_4,\nonumber\\
g_0^{(21)}=&-\frac{1}{15}
[G']_{13},\nonumber\\
g_0^{(22)}=&\frac{1}{5}
[G']_{14},\nonumber\\
g_2^{(11)}=&-\frac{5}{21}
[G']_1+\frac{5}{7}[G']_2,\nonumber\\
g_2^{(12)}=&-\frac{4}{7}
[G']_3
+
\frac{4}{21}
[G']_4,\nonumber\\
g_2^{(21)}=&-\frac{4}{7}
[G']_{11}+
\frac{4}{21}
[G']_{13},\nonumber\\
g_2^{(22)}=&\frac{6}{7}
[G']_{12}-
\frac{2}{7}
[G']_{14},\nonumber\\
g_2^{(31)}=&-\frac{4}{7}
[G']_7+
\frac{4}{21}
[G']_{13},\nonumber\\
g_2^{(32)}=&\frac{6}{7}
[G']_8-
\frac{2}{7}
[G']_{14},\nonumber\\
g_2^{(41)}=&-\frac{5}{21}
[G']_1+
\frac{5}{7}
[G']_5,\nonumber\\
g_2^{(42)}=&\frac{4}{21}
[G']_4-
\frac{4}{7}
[G']_{10},\nonumber\\
g_4^{(11)}=&
-\frac{153}{245}[G']_1
+\frac{8}{7}[G']_2
-\frac{12}{7}[G']_3
+\frac{184}{245}[G']_4
+\frac{8}{7}[G']_5\nonumber\\
&-\frac{12}{7}[G']_7
+\frac{4}{7}[G']_8
+4[G']_9
-\frac{12}{7}[G']_{10}
-\frac{12}{7}[G']_{11}\nonumber\\
&+\frac{4}{7}[G']_{12}
+\frac{184}{245}[G']_{13}
-\frac{122}{245}[G']_{14}.
\end{align}

\subsection{Isotropic average of electric quadrupole term}
The rotational average of electric quadrupole contribution requires averaging of a ninth-rank tensors rather than eighth-rank tensors for the case of electric and magnetic dipole contributions. However, rotational average of ninth- and eleventh-rank tensors can be found in our previous work \cite{Begzjav2019}.
Using the result of our paper \cite{Begzjav2019}, we obtain the rotational average of last four terms in Eq.~\ref{general_vvvr} as follows
\begin{align}\label{averageelectricquarupole1}
&-\frac{k_3}{3}
\langle
A_{y,xz}(-\omega_3,\omega_4)
\alpha_{xx}(-\omega_1,\omega_2)
\alpha_{xx}(-\omega_3,\omega_4)
\alpha_{xx}(-\omega_1,\omega_2)\rangle\nonumber\\
&+\frac{k_3}{3}
\langle
A_{x,xz}(-\omega_3,\omega_4)
\alpha_{xx}(-\omega_1,\omega_2)
\alpha_{yx}(-\omega_3,\omega_4)
\alpha_{xx}(-\omega_1,\omega_2)\rangle\nonumber\\
&+
\frac{k_4}{3}
\langle
A_{x,yz}(-\omega_3,\omega_4)
\alpha_{xx}(-\omega_1,\omega_2)
\alpha_{xx}(-\omega_3,\omega_4)
\alpha_{xx}(-\omega_1,\omega_2)\rangle\nonumber\\
&-\frac{k_4}{3}
\langle
A_{x,xz}(-\omega_3,\omega_4)
\alpha_{xx}(-\omega_1,\omega_2)
\alpha_{yx}(-\omega_3,\omega_4)
\alpha_{xx}(-\omega_1,\omega_2)\rangle\nonumber\\
&=
-
\frac{k_3}{3}\left(
\frac{48}{22680}[A]_5
+\frac{96}{22680}[A]_6
+\frac{96}{22680}[A]_7
+\frac{192}{22680}[A]_8\right.\nonumber\\
&+\frac{192}{22680}[A]_9
+\frac{96}{22680}[A]_{10}
+\frac{144}{22680}[A]_{11}
+\frac{288}{22680}[A]_{12}\nonumber\\
&+\left.\frac{36}{22680}[A]_{13}
+\frac{72}{22680}[A]_{14}\right)\nonumber\\
&+
\frac{k_4}{3}\left(
\frac{48}{22680}[A]_{11}
+\frac{96}{22680}[A]_{12}
+\frac{12}{22680}[A]_{13}
+\frac{24}{22680}[A]_{14}\right),
\end{align}
where
\begin{widetext}
\begin{align}
[A]_5&=
\varepsilon_{mni}
A_{m,nj}(-\omega_3,\omega_4)
\alpha_{ij}(-\omega_1,\omega_2)
\alpha_{kk}(-\omega_3,\omega_4)
\alpha_{ll}(-\omega_1,\omega_2),\nonumber\\
[A]_6&=
\varepsilon_{mni}
A_{m,nj}(-\omega_3,\omega_4)
\alpha_{ij}(-\omega_1,\omega_2)
\alpha_{kl}(-\omega_3,\omega_4)
\alpha_{kl}(-\omega_1,\omega_2),\nonumber\\
[A]_7&=
\varepsilon_{mni}
A_{m,nj}(-\omega_3,\omega_4)
\alpha_{ik}(-\omega_1,\omega_2)
\alpha_{jk}(-\omega_3,\omega_4)
\alpha_{ll}(-\omega_1,\omega_2),\nonumber\\
[A]_8&=
\varepsilon_{mni}
A_{m,nj}(-\omega_3,\omega_4)
\alpha_{ik}(-\omega_1,\omega_2)
\alpha_{jl}(-\omega_3,\omega_4)
\alpha_{kl}(-\omega_1,\omega_2),\nonumber\\
[A]_9&=
\varepsilon_{mni}
A_{m,nj}(-\omega_3,\omega_4)
\alpha_{ik}(-\omega_1,\omega_2)
\alpha_{kl}(-\omega_3,\omega_4)
\alpha_{jl}(-\omega_1,\omega_2),\nonumber\\
[A]_{10}&=
\varepsilon_{mni}
A_{m,nj}(-\omega_3,\omega_4)
\alpha_{ik}(-\omega_1,\omega_2)
\alpha_{ll}(-\omega_3,\omega_4)
\alpha_{jk}(-\omega_1,\omega_2),\nonumber\\
[A]_{11}&=
\varepsilon_{mni}
A_{m,nj}(-\omega_3,\omega_4)
\alpha_{jk}(-\omega_1,\omega_2)
\alpha_{ik}(-\omega_3,\omega_4)
\alpha_{ll}(-\omega_1,\omega_2),\nonumber\\
[A]_{12}&=
\varepsilon_{mni}
A_{m,nj}(-\omega_3,\omega_4)
\alpha_{jk}(-\omega_1,\omega_2)
\alpha_{il}(-\omega_3,\omega_4)
\alpha_{kl}(-\omega_1,\omega_2),\nonumber\\
[A]_{13}&=
\varepsilon_{mni}
A_{m,nj}(-\omega_3,\omega_4)
\alpha_{kk}(-\omega_1,\omega_2)
\alpha_{ij}(-\omega_3,\omega_4)
\alpha_{ll}(-\omega_1,\omega_2),\nonumber\\
[A]_{14}&=
\varepsilon_{mni}
A_{m,nj}(-\omega_3,\omega_4)
\alpha_{kl}(-\omega_1,\omega_2)
\alpha_{ij}(-\omega_3,\omega_4)
\alpha_{kl}(-\omega_1,\omega_2).
\end{align}
\end{widetext}
Here, $\varepsilon_{mni}$ is Levi-Civita symbol that is anti-symmetric third rank tensor.
It is obvious that there is no $[A]_1,\ldots,[A]_4$ terms since contraction $\varepsilon_{mni}A_{m,ni}$ vanishes due to symmetry properties of tensor $A_{m,ni}$.
As previous cases, the isotropic invariants $[A]_i$ are, again, not independent. Connecting relation is given by
\begin{align}
&-2[A]_5
+2[A]_6
+2[A]_7
-2[A]_8
-2[A]_9
+2[A]_{10}
+2[A]_{11}\nonumber\\
&-2[A]_{12}
-[A]_{13}
+[A]_{14}
=0.
\end{align}
This relation is electric quadrupole analog of relation \ref{relationmagneticdipole}, and only difference is the absence of $[A]_1\ldots [A]_4$ terms.
In terms of natural invariants $k_J^{(\tau_1\tau_2)}$, Eq.~\ref{averageelectricquarupole1} is as follows:
\begin{align}
&-\frac{k_3}{3}
\langle
A_{y,xz}(-\omega_3,\omega_4)
\alpha_{xx}(-\omega_1,\omega_2)
\alpha_{xx}(-\omega_3,\omega_4)
\alpha_{xx}(-\omega_1,\omega_2)\rangle\nonumber\\
&+\frac{k_3}{3}
\langle
A_{x,xz}(-\omega_3,\omega_4)
\alpha_{xx}(-\omega_1,\omega_2)
\alpha_{yx}(-\omega_3,\omega_4)
\alpha_{xx}(-\omega_1,\omega_2)\rangle\nonumber\\
&+
\frac{k_4}{3}
\langle
A_{x,yz}(-\omega_3,\omega_4)
\alpha_{xx}(-\omega_1,\omega_2)
\alpha_{xx}(-\omega_3,\omega_4)
\alpha_{xx}(-\omega_1,\omega_2)\rangle\nonumber\\
&-\frac{k_4}{3}
\langle
A_{x,xz}(-\omega_3,\omega_4)
\alpha_{xx}(-\omega_1,\omega_2)
\alpha_{yx}(-\omega_3,\omega_4)
\alpha_{xx}(-\omega_1,\omega_2)\rangle\nonumber\\
&=
\frac{1}{3c}\left(
+\frac{7853}{92610}k_{0,\omega_3}^{(21)}
-\frac{7853}{138915}k_{0,\omega_3}^{(22)}
+\frac{5}{378}k_{2,\omega_3}^{(21)}
-\frac{10}{567}k_{2,\omega_3}^{(22)}\right.\nonumber\\
&+\left.\frac{1}{105}k_{2,\omega_3}^{(31)}
-\frac{4}{315}k_{2,\omega_3}^{(32)}
-\frac{2}{525}k_{2,\omega_3}^{(41)}
+\frac{1}{105}k_{2,\omega_3}^{(42)}
-\frac{1}{315}k_{4,\omega_3}^{(11)}
\right]\nonumber\\
&
+\frac{1}{3c}
\left(
\frac{1}{54}k_{0,\omega_4}^{(21)}
-\frac{1}{81}k_{0,\omega_4}^{(22)}
+\frac{1}{270}k_{2,\omega_4}^{(21)}
-\frac{2}{405}k_{2,\omega_4}^{(22)}
\right).
\end{align}
Since tensor product $\varepsilon_{mni}A_{m,nj}$ transforms like second-rank tensor $G'_{ij}$ under the three dimensional rotation, we define natural invariants of tensor products $A_{m,on}(-\omega_3,\omega_4)
\alpha_{pq}(-\omega_1,\omega_2)
\alpha_{ij}(-\omega_3,\omega_4)
\alpha_{kl}(-\omega_1,\omega_2)$ as definition of $g_j^{(\tau_1\tau_2)}$ given in previous subsection. There is no difference except vanishing terms $[A]_1,\ldots,[A]_4$. These are
\begin{align}
k_{0,\omega_i}^{(21)}=&-\frac{\omega_i}{15}
[A]_{13},\nonumber\\
k_{0,\omega_i}^{(22)}=&\frac{\omega_i}{5}
[A]_{14},\nonumber\\
k_{2,\omega_i}^{(21)}=&-\frac{4\omega_i}{7}
[A]_{11}+
\frac{4\omega_i}{21}
[A]_{13},\nonumber\\
k_{2,\omega_i}^{(22)}=&\frac{6\omega_i}{7}
[A]_{12}-
\frac{2\omega_i}{7}
[A]_{14},\nonumber\\
k_{2,\omega_i}^{(31)}=&-\frac{4\omega_i}{7}
[A]_7+
\frac{4\omega_i}{21}
[A]_{13},\nonumber\\
k_{2,\omega_i}^{(32)}=&\frac{6\omega_i}{7}
[A]_8-
\frac{2\omega_i}{7}
[A]_{14},\nonumber\\
k_{2,\omega_i}^{(41)}=&
\frac{5\omega_i}{7}
[A]_5,\nonumber\\
k_{2,\omega_i}^{(42)}=&-
\frac{4\omega_i}{7}
[A]_{10},\nonumber\\
k_{4,\omega_i}^{(11)}=&
\frac{8\omega_i}{7}[A]_5
-\frac{12\omega_i}{7}[A]_7
+\frac{4\omega_i}{7}[A]_8\nonumber\\
&+4\omega_i[A]_9
-\frac{12\omega_i}{7}[A]_{10}
-\frac{12\omega_i}{7}[A]_{11}
+\frac{4\omega_i}{7}[A]_{12}\nonumber\\
&+\frac{184\omega_i}{245}[A]_{13}
-\frac{122\omega_i}{245}[A]_{14}.
\end{align}
As in the theory of Raman optical activity, the definition of natural invariants $k_j^{(\tau_1,\tau_2)}$ incorporates the frequencies $\omega_3$ and $\omega_4$.

\bibliography{aipsamp}

\end{document}